# Subwavelength Coherent Scaling of High-Order Nonlinear Light Generation in Bulk Monolayer $MoS_2$ Thin Films

Boxuan Zhou[1†], Yuancheng Jing[2†], Chun-Chieh Yu[2†], Haoyang Li[1], Ran Wang[1], Xingxu Yan[3], Xiaoqing Pan[3,4], Yu Huang[1,5,6], Wei Xiong[2*], Xiangfeng Duan[1,6*]

[1]Department of Chemistry and Biochemistry, University of California, Los Angeles, Los Angeles, USA.

[2]Department of Chemistry and Biochemistry, University of California, San Diego, San Diego, USA.

[3]Department of Materials Science and Engineering, University of California, Irvine, Irvine, USA.

[4]Department of Physics and Astronomy, University of California, Irvine, Irvine, USA.

[5]Department of Materials Science and Engineering, University of California, Los Angeles, Los Angeles, USA.

[6]California NanoSystems Institute, University of California, Los Angeles, Los Angeles, USA.

[†]These authors contributed equally: Boxuan Zhou, Yuancheng Jing, Chun-Chieh Yu

*Corresponding authors. Email: xduan@chem.ucla.edu; w2xiong@ucsd.edu

**Abstract:** Monolayer transition metal dichalcogenides (e.g., $MoS_2$) exhibit exceptionally large optical nonlinearities for high-order nonlinear light generation (NLG), yet their inherent atomic thickness fundamentally limits light-matter interactions and thus conversion efficiency. Here, we overcome this intrinsic trade-off using a solution-processed bulk monolayer $MoS_2$ (BM-$MoS_2$) architecture composed of electronically decoupled $MoS_2$ monolayers separated by organic interlayers. This layered superstructure preserves the exceptional intrinsic nonlinear susceptibility of monolayer $MoS_2$ while enabling scalable interaction length. In the subwavelength regime, the NLG scales nearly quadratically with layer number ($N^{1.8}$), confirming the constructive buildup of nonlinear fields across stacked monolayers. As a result, a 100-nm-thick BM-$MoS_2$ thin film exhibits colossal high-order NLG, including four-wave mixing and high-harmonic generation. The generated nonlinear beam is directly visible to the naked eye and exhibits broad spectral tunability spanning approximately 1000 nm in the mid-IR, enabling mid-IR-to-visible upconversion spectroscopy for resolving molecular vibrational fingerprints. By uniting monolayer-scale nonlinear susceptibility with bulk interaction length and coherent field buildup, BM-$MoS_2$ establishes a thin-film platform for ultra-compact and substrate-agnostic nonlinear photonic systems beyond the constraints of conventional single crystals.

## INTRODUCTION

Nonlinear light generation (NLG) underpins coherent frequency conversion and is central to modern photonics, enabling applications ranging from ultrafast spectroscopy to quantum information processing (*1-9*). Beyond the prototypical second harmonic generation (SHG), high-order nonlinear processes such as multi-beam wave mixing and high-harmonic generation (HHG) can extend frequency conversion across a much broader spectral range from the terahertz to the X-ray domain, and offer pathways to photonic computation and ultra-fast laser generation and manipulation (*1, 8-22*).

Conventional NLG relies on **macroscopic bulk crystals** grown at high temperatures (often >500 °C) (*23-26*), where long interaction lengths are required to compensate for intrinsically weak nonlinear susceptibilities. However, this introduces **a fundamental limitation: as thickness increases beyond the coherence length, typically ranging from a few hundred nanometers to micrometers, destructive interference arising from phase mismatch suppresses nonlinear conversion,** particularly for multi-beam wave mixing and HHG involving large frequency separations. Nanophotonic resonators and metasurfaces can partially alleviate this phase-matching constraint through field confinement and amplification (*6, 10, 27*), but their operation is inherently narrowband (typically <30 nm) and demands complex nanofabrication, restricting scalability and integration with flexible or low-thermal-budget photonic platforms.

Thin-film nonlinear materials offer an alternative route by operating in the subwavelength thickness regime (*28-31*). Although phase-matching constraints are naturally relaxed, their performance is fundamentally limited by the short interaction length, necessitating exceptionally large intrinsic nonlinear susceptibilities. As a result, **most existing thin-film nonlinear systems exhibit conversion efficiencies that remain too low for practical applications.**

Monolayer transition-metal dichalcogenides (TMDs) have emerged as promising candidates owing to their exceptionally large nonlinear susceptibilities arising from strong excitonic effects in the atomically thin limit. These properties enable enhanced light–matter interactions and efficient high-order nonlinear responses. However, their sub-nanometer thickness fundamentally restricts interaction length. As a result, the practical conversion efficiencies remain extremely low for higher-order nonlinear processes. For example, the conversion efficiencies of third harmonic generation (THG) and four-wave mixing (FWM) are typically on the order of $10^{-8}$ % to $10^{-6}$ % (Table S1) (*32-36*). **Increasing thickness into multilayer or bulk TMDs defeats the intrinsic merits of monolayer TMDs**, as interlayer coupling, dielectric screening, and symmetry restoration rapidly suppress exciton-enhanced nonlinearities and completely quench even-order processes (*22, 34, 37, 38*). **These limitations highlight a critical challenge in implementing TMD materials for nonlinear optics: preserving the exceptional nonlinear susceptibility of monolayer crystals while achieving scalable interaction lengths.**

Here we resolve this long-standing trade-off by introducing a solution-processed bulk monolayer $MoS_2$ (BM-$MoS_2$) architecture—a quasi-two-dimensional (2D) superlattice of electronically decoupled $MoS_2$ monolayers separated by molecular spacers. This design preserves the giant intrinsic nonlinear susceptibility of monolayer $MoS_2$ while enabling thickness-scalable interaction length. In the subwavelength regime, the nonlinear response exhibits near-quadratic scaling with thickness, indicating coherent accumulation of nonlinear polarization across stacked monolayers within a sub-coherence-length regime. As a result, BM-$MoS_2$ thin films exhibit colossal multi-order NLG, including four-wave mixing (FWM), sum-frequency generation (SFG), and HHG up to the fifth order over a broad spectral range. A 100-nm-thick film can generate nonlinear beams directly visible to the naked eye. BM-$MoS_2$ thin films further enable mid-infrared-to-visible upconversion spectroscopy capable of resolving molecular vibrational fingerprints. By uniting monolayer-scale nonlinear susceptibility with bulk-scale interaction length and coherent polarization buildup, BM-$MoS_2$ establishes an integrable, low-cost, and high-performance nonlinear thin film material platform beyond the limitations of conventional bulk crystals.

## RESULTS

### BM-$MoS_2$ preparation and characterization

Monolayer $MoS_2$ ink was first prepared through an electrochemical molecular intercalation and exfoliation approach and stabilized as a colloidal dispersion capped with surface molecular ligands (photograph in Fig. 1D; see Methods for details), and then solution-assembled into solid-state thin films composed of alternating monolayer $MoS_2$ crystals and organic molecular spacers (Fig. 1A and B). The interleaving insulating and low-dielectric-constant organic molecular spacer effectively decouples interlayer electronic interactions among the $MoS_2$ layers in the multilayered thin films, effectively preserving monolayer optical properties and achieving thickness-scalable optical length for versatile NLG (Fig. 1B and C). The atomic force microscopy (AFM) image reveals a uniform distribution of the exfoliated $MoS_2$ nanosheets in the colloidal ink (Fig. 1E). High-resolution AFM further showed that the nanosheet has a thickness of around 2.5 nm (Fig. 1F), which is much larger than that of pristine monolayer $MoS_2$ (~0.62 nm) due to the presence of polyvinylpyrrolidone (PVP) ligand layers adsorbed on both sides of the monolayer $MoS_2$. The PVP encapsulation is also evidenced by the spatially resolved electron energy loss spectroscopy (EELS) elemental mapping (Fig. S1). Subsequently, nanosheet thin films were fabricated by spray-coating the ink onto silicon substrates. The top-view scanning electron microscopy (SEM) image shows a continuous film composed of staggered nanosheets (Fig. 1G, upper), while the cross-sectional SEM and AFM measurements reveal a film thickness of ~100 nm (Fig. 1G, bottom; Fig. S2). The film thickness can be tailored by adjusting the ink-coating parameters (see Methods for details). The X-ray diffraction (XRD) analysis displayed an interlayer periodicity of ~2.6 nm (red curve, Fig. 1H), which agrees well with the AFM-measured thickness of a single $MoS_2$ nanosheet and considerably larger than the interlayer distance of pristine bulk $MoS_2$ (0.62 nm, black curve in Fig. 1H). These results confirm the formation of an ordered, alternating PVP/$MoS_2$/PVP superlattice thin film structure.

The PVP interlayer spacer (~2 nm thick) possesses a large HOMO-LUMO gap (>4.4 eV, insulating) and a low dielectric constant (~2.2), which together suppress interlayer electronic coupling and minimize dielectric screening. As a result, intrinsic monolayer excitonic characteristics are well preserved throughout the entire film, which can be referred to as BM-$MoS_2$. **It is well-known that interlayer coupling in $MoS_2$ results in a direct-to-indirect bandgap transition (*39*), as evidenced by the quenched photoluminescence (PL) in pristine bulk $MoS_2$ (black curve, Fig. 1I). In contrast, the BM-$MoS_2$ film displays prominent A-exciton luminescence resembling that of pristine monolayer $MoS_2$, indicating the preservation of direct-gap excitonic properties** (red curve, Fig. 1I). The wavenumber difference between the in-plane ($E_{2g}^1$) and out-of-plane ($A_{1g}$) Raman modes in BM-$MoS_2$ is also consistent with that of pristine monolayer $MoS_2$ (19.2 $cm^{-1}$) and notably smaller than that of bulk $MoS_2$ (25.2 $cm^{-1}$), confirming the absence of interlayer coupling (Fig. 1J) (*40*). We note that BM-$MoS_2$ thin film preparations were carried out at room temperature under ambient conditions, establishing a versatile solution processable thin film material can be readily integrated on arbitrary substrate, including polymers and non-planar surfaces.

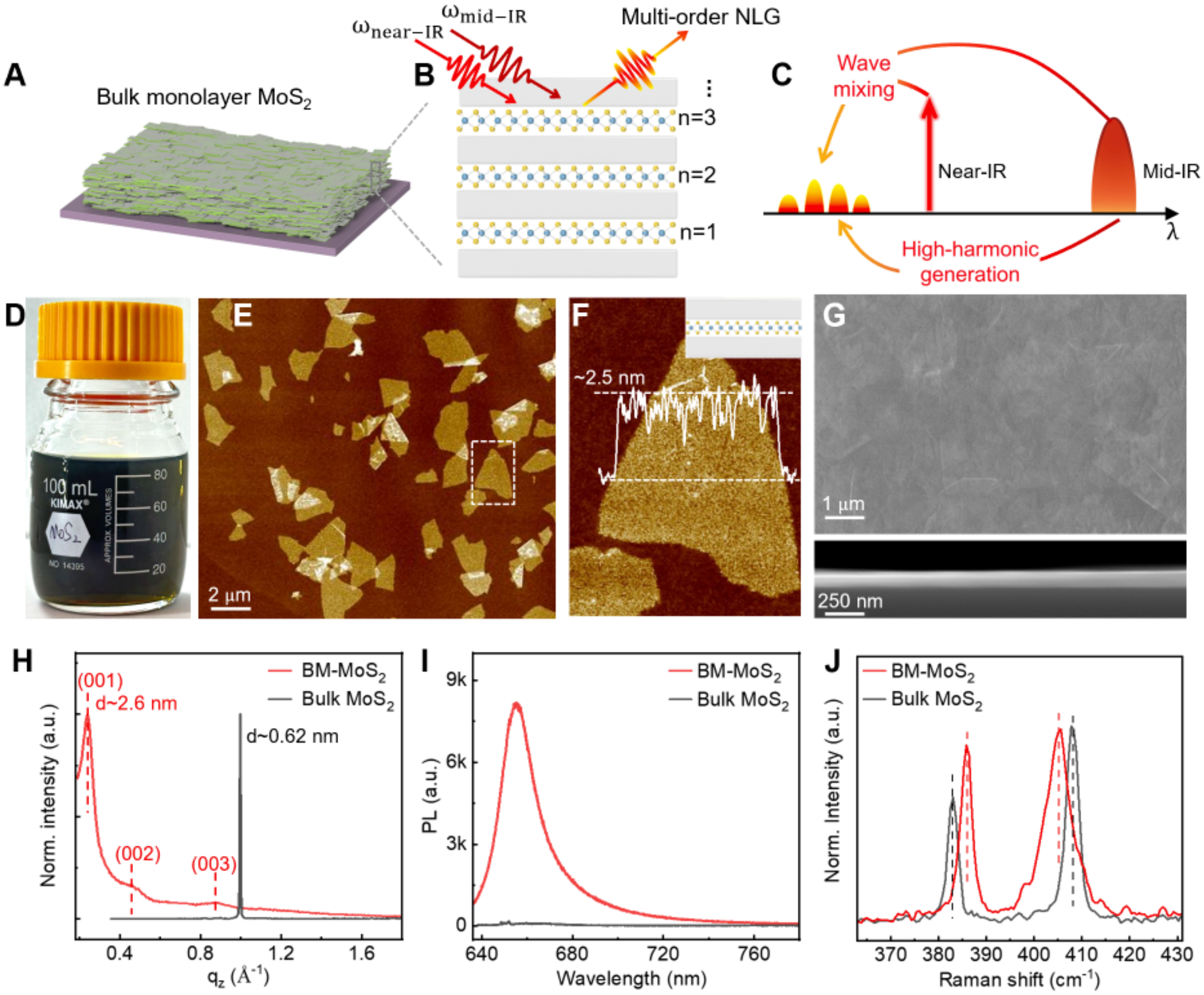

**Fig. 1. Preparation of bulk monolayer $MoS_2$ (BM-$MoS_2$) for nonlinear light generation.** (**A**) Schematics of BM-$MoS_2$ thin films. (**B** and **C**) Schematic illustration of multi-order harmonic generation and wave mixing from a multilayered BM-$MoS_2$ excited by near-IR and mid-IR light. The n value in (B) indicates the number of PVP/$MoS_2$-monolayer /PVP repeating units. (**D**) Photograph of the monolayer $MoS_2$ colloidal ink. (**E** and **F**) Atomic force microscopy (AFM) images of monolayer $MoS_2$ nanosheets from the colloidal ink (E), showing uniform distribution of the as-exfoliated $MoS_2$ flakes. Each nanosheet is encapsulated by PVP (see the zoomed-in AFM image in F), with an average flake thickness of around 2.5 nm. (**G**) Top view and cross-sectional scanning electron microscopy (SEM) images of a BM-$MoS_2$ thin film deposited on a silicon wafer. (**H** to **J**) XRD (H), PL (I), and Raman (J) spectra of the pristine bulk $MoS_2$ (black curves) and BM-$MoS_2$ (red curves).

## Multi-order NLG in BM-$MoS_2$

NLG measurements were performed under ambient conditions using a 45° reflection geometry (Fig. 2A). A 100 nm-thick BM-$MoS_2$ thin film was excited by two collinearly aligned laser beams: a near-IR laser beam with a fixed wavelength of 1028.9 nm (abbreviated by its angular frequency $\omega_1$) and a mid-IR laser beam with tunable wavelengths (abbreviated by $\omega_2$) (see Methods for details). Strikingly, the BM-$MoS_2$ thin films produce visible emission bright enough to be seen with the naked eye (Fig. 2A), providing direct evidence of strong nonlinear conversion. Remarkably, this nonlinear light maintains a well-defined beam mode as it propagates over meter-scale distances in free-space (Fig. 2A, Fig. S3).

Spectral analysis reveals six distinct NLG processes in the BM-$MoS_2$ under two-color excitation within our detector spectral range (Fig. 2B), as illustrated in the energy-level diagrams (Fig. 2C). The strongest emissions arise from FWM processes that generate idler frequencies at $2\omega_1 - \omega_2$ and $\omega_1 + 2\omega_2$, and a 5$^{th}$HG of the mid-IR driving laser ($5\omega_2$) (red spectrum in Fig. 2B). We also observed several even-order nonlinear signals including SHG of the near-IR driving laser ($2\omega_1$) (green spectra), sum-frequency generation (SFG, $\omega_1$+$\omega_2$), and the 4$^{th}$ harmonic of the mid-IR ($4\omega_2$) (black spectra). Notably, the peak intensities of SFG and 4$^{th}$HG are over four orders of magnitude lower than those of their higher-photon-energy odd-order counterparts (e.g., SFG vs. FWM; 4$^{th}$HG vs. 5$^{th}$HG). We note that the SHG, SFG, and 4$^{th}$HG signals are generated at wavelengths far apart from the exciton-enhanced spectral window of $MoS_2$, which likely account for their much weaker intensities compared with the dominant FWM and 5$^{th}$HG signals. Under our excitation conditions, the SHG signal from the near-IR pump appears at ~514.5 nm, while the SFG signal appears near ~780 nm. Both wavelengths fall outside the 600-650 nm spectral window where the strongest exciton-enhanced FWM and 5$^{th}$HG responses are observed (see details in the later section). Additionally, the different polarization characteristics of even and odd nonlinear processes can also contribute to their intensity disparity, as discussed in a later section.

**The substantial intensity difference among these nonlinear processes also rules out cascaded emission pathways and confirms that the observed higher order NLG originates from direct nonlinear interactions in the BM-$MoS_2$**. For example, the $\omega_1 + 2\omega_2$ emission is a direct FWM process and cannot be produced by cascaded process of two SFG steps. Likewise, the bright 5$^{th}$ HG ($5\omega_2$) signal cannot result from the 4$^{th}$ HG that subsequently mixes with $\omega_2$. The dominance of these direct multi-order nonlinear interactions reflects the intrinsic nonlinear response of the BM-$MoS_2$ architecture.

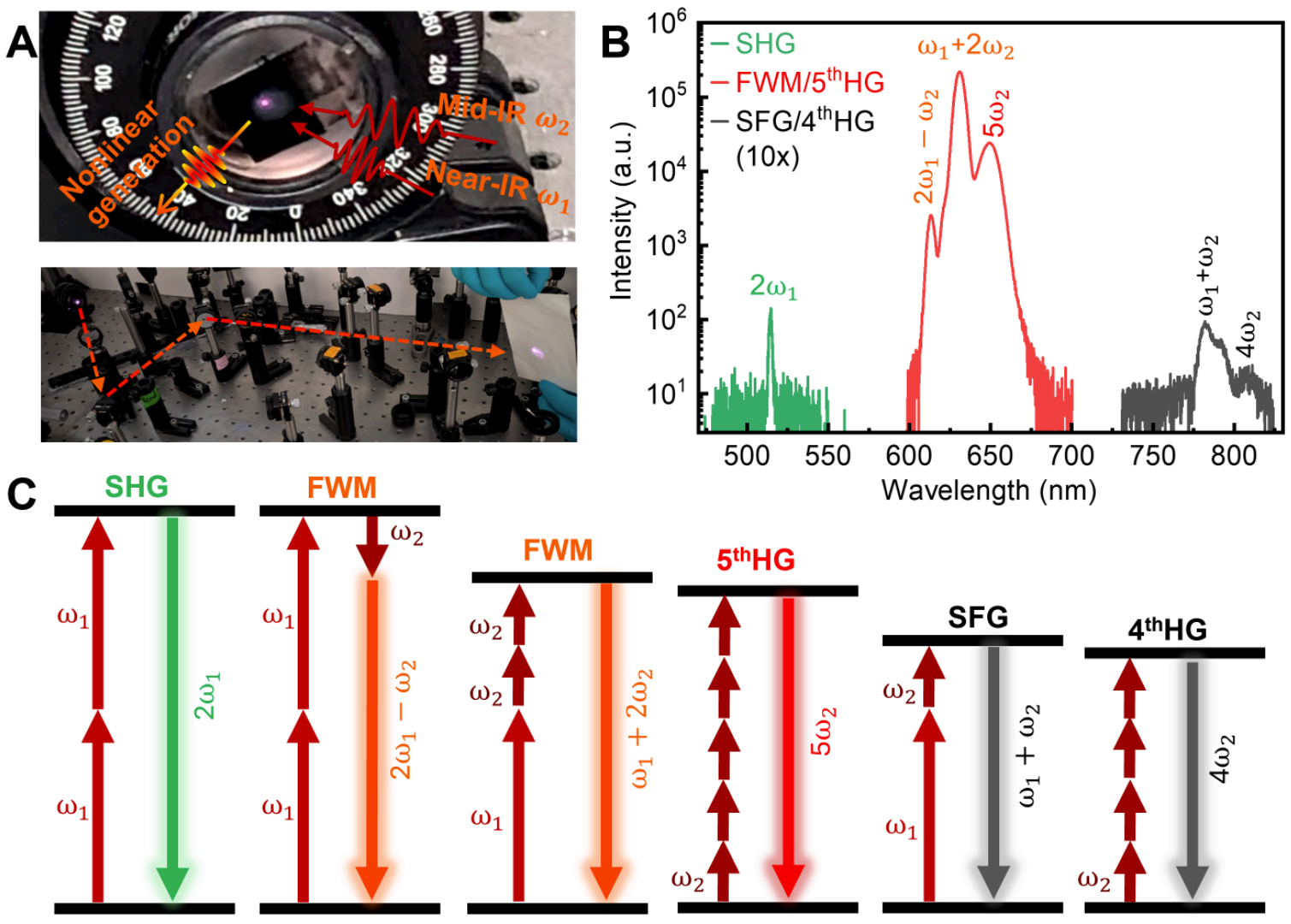

**Fig. 2. Versatile nonlinear light generation in BM-$MoS_2$ thin films.** (**A**) Photographs of the measurement setup using a 45° reflection measurement geometry under near-IR and mid-IR dual color excitation. The Near-IR laser (abbreviated by $\omega_1$) has a fixed wavelength of 1028.9 nm and the mid-IR laser (abbreviated by $\omega_2$) has tunable wavelengths. The NLG spot is visibly observed on the BM-$MoS_2$ film (top) and is reflected at 135°, propagating over meter-scale distances in free space (bottom). The dashed arrows indicate the beam paths. (**B**) Emission spectra of a BM-$MoS_2$ thin film under dual-color ($\omega_1$ and $\omega_2$) excitation, showing distinct peaks corresponding to second harmonic generation ($2\omega_1$, green curve), four-wave mixing processes (FWM, $2\omega_1$-$\omega_2$ and $\omega_1$+$2\omega_2$, red curve), 5$^{th}$ harmonic generation (5$^{th}$HG, $5\omega_2$, red curve), sum frequency generation ($\omega_1$+$\omega_2$, black curve), and 4$^{th}$ harmonic generation (4$^{th}$HG, $4\omega_2$, black curve). (**C**) Schematics of the energy level diagrams of these NLG processes.

## FWM and 5$^{th}$HG in BM-$MoS_2$

We next focus our discussions on the FWM and 5$^{th}$HG, which is confirmed by comparing the emission spectra under single-color ($\omega_2$ only) and two-color ($\omega_1$ and $\omega_2$) excitation. When the near-IR field $\omega_1$ is removed, all FWM signals disappear completely, while the 5$^{th}$HG of the mid-IR field ($\omega_2$) remains unchanged (Fig. 3A). The invariance of the 5$^{th}$HG upon removal of $\omega_1$ confirms that it is solely driven by the mid-IR $\omega_2$ field. We next scanned the wavelength of the mid-IR pump ($\omega_2$) from 2470 nm to 3470 nm. The resulting spectra clearly track the expected tuning behaviors of the two processes. Under two-color excitation, the main FWM idler peak shifts according to $\omega_1$+$2\omega_2$, while the 5$^{th}$HG spectra collected under single-color excitation follow the scaling of $5\omega_2$. Both nonlinear emissions exhibit continuous and monotonic frequency shifts across the entire tuning range, confirming their coherent nonlinear origins and the broad spectral tunability (Fig. 3B and C). We note that nonlinear emission spectra corresponding to $\omega_2$ wavelengths overlapping with the water-vapor absorption band are omitted, as strong atmospheric absorption at these wavelengths leads to a substantial attenuation of the $\omega_2$ pump power.

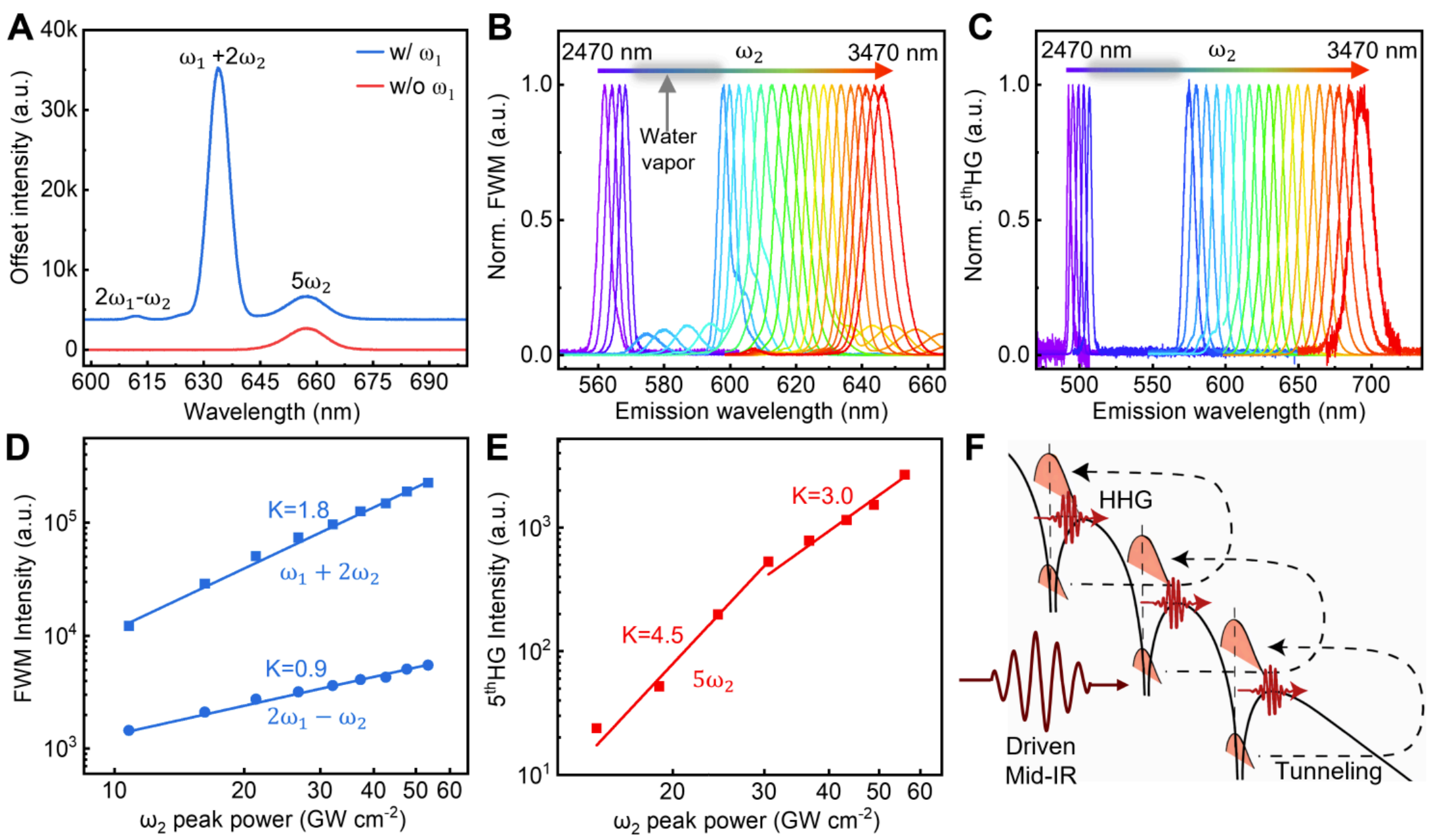

**Fig. 3**. **FWM and 5$^{th}$HG in BM-$MoS_2$.** (**A**) Emission spectra of the BM-$MoS_2$ thin film under dual color ($\omega_1$ and $\omega_2$, top) and single color ($\omega_2$, bottom) excitation. (**B** and **C**) Normalized nonlinear emission spectra from the BM-$MoS_2$ thin film under two-color (B) and single-color (C) with tunable $\omega_2$ (2470 nm to 3470 nm). $\omega_1$ laser wavelength is fixed at 1028.9 nm. Spectra acquired at $\omega_2$ wavelengths overlapping the atmospheric water-vapor absorption band (grey region on the arrow) are omitted because strong absorption substantially attenuates the pump power. (**D** and **E**) Two FWM emission channels (D) and 5$^{th}$HG (E) as a function of $\omega_2$ peak power density. Solid lines are power-law fits ($I \propto P^K$) of FWM and 5$^{th}$HG peak intensity as a function of $\omega_2$ peak power density. (**F**) Schematics of non-perturbative nonlinear light emission (*17*). In the non-perturbative regime, the mid-IR driving field is strong enough to create electron-hole pairs through electron tunneling across the bandgap, followed by carrier acceleration and electron-hole recollision that result in high-harmonic photon radiation (*17*).

We also conducted excitation power dependent measurements to verify the nonlinear origins of FWM and 5$^{th}$HG. For the FWM process $\omega_{FWM} = \omega_1 + 2\omega_2$, the generated field follows $E_{FWM} \propto \chi^{(3)} E_{\omega_1} E_{\omega_2} E_{\omega_2}$, where $E_{\omega_1}$ and $E_{\omega_2}$ denote the electric field amplitudes of the near-IR and mid-IR pump beams, respectively. Accordingly, the FWM peak intensity is expected to scale quadratically with the $\omega_2$ pump power. For the FWM process $\omega_{FWM} = 2\omega_1 - \omega_2$, the relation $E_{FWM} \propto \chi^{(3)} E_{\omega_1} E_{\omega_1} E_{\omega_2}$ predicts a linear power dependence on the $\omega_2$ pump power. These scaling behaviors are experimentally confirmed by varying the excitation power (Fig. 3D).

The excitation-power-dependence of the 5$^{th}$HG reveals an additional aspect of the nonlinear response. At moderate excitation, the 5$^{th}$HG output follows a power exponent (K) of 4.5 when the peak power density is below 30 GW cm$^{-2}$ (Fig. 3E). This behavior is close to the expected fifth-order scaling (K=5) in the perturbative regime, where the process can be described by a multi-photon absorption picture (Fig. 2C). At higher excitation powers, the 5$^{th}$HG deviates markedly from the perturbative behavior, showing the power exponent decreasing to ~3.0. Previous studies on pristine monolayer $MoS_2$ have reported a similar transition in high-order harmonics (order>4), where **the power exponent approaches a universal value near 3 in the non-perturbative regime**. Our results therefore indicate that the 5$^{th}$HG in BM-$MoS_2$ enters the **non-perturbative regime, where the HHG originate from interband tunneling and intraband carrier acceleration triggered by the intense mid-IR field (Fig. 3F) (*12, 17*).** We note that the BM-$MoS_2$ thin film remains stable throughout our measurements, and the observed power-law transition arises from the intrinsic nonlinear response rather than sample degradation.

**Colossal FWM and 5$^{th}$HG in BM-$MoS_2$**

To evaluate the FWM and 5$^{th}$HG performance of BM-$MoS_2$ thin films, we have benchmarked against a standard ZnSe nonlinear crystal (3 mm thick), a nonlinear crystal with well-established $\chi^{(3)}$ and HHG performance. Remarkably, **the FWM emissions at $\boldsymbol{\omega_1 + 2\omega_2}$ and $\boldsymbol{2\omega_1 - \omega_2}$ in the BM-$MoS_2$ are 85 and 66 times stronger than those from ZnSe, respectively (red vs. black spectra in Fig. 4A), despite the 100-nm-thick BM-$MoS_2$ thin film being four orders of magnitude thinner**. **The 5$^{th}$HG is even more striking, with BM-$MoS_2$ exhibiting a 180-fold stronger 5$^{th}$HG than ZnSe (Fig. 4A and Fig. S4). Additionally, FWM and 5$^{th}$HG intensities of the BM-$MoS_2$ thin film exceed those of the pristine 2H-phase $MoS_2$ bulk single crystal by factors of 19 and 125, respectively (red vs. blue spectra in Fig. 4A and Fig. S4).**

A key distinction of BM-$MoS_2$ relative to pristine bulk $MoS_2$ and ZnSe is its monolayer-like exciton resonances. To assess the excitonic contribution to the colossal nonlinear responses, we performed wavelength-dependent excitation spectroscopy in BM-$MoS_2$. As shown in Fig. 4B, the extracted FWM intensities plotted versus emission wavelength exhibit a pronounced resonance enhancement centered at ~625 nm, corresponding to a ~ 30 nm (90 meV) blueshift relative to the A-exciton peak in the linear optical absorption spectrum (Fig. S5). Such a blueshift is consistent with the previous THG excitation spectroscopy and pump-probe studies of TMDs driven by intense mid-IR field (>10 GW $cm^{-2}$), and are attributed to **strong-field exciton renormalization** (light-dressing effect) (*33, 41*). In this picture, the intense driving field imparts ponderomotive energy and mixes bound exciton states with the quasi-free continuum, shifting the effective excitonic energy toward higher levels (*41*). The extracted $5^{th}$HG intensities exhibit a similar blueshift, though with a considerably broader dispersion than FWM (Fig. 4C), reflecting a more complex involvement of strong-field-dressed excitonic effects in higher-order harmonic generation.

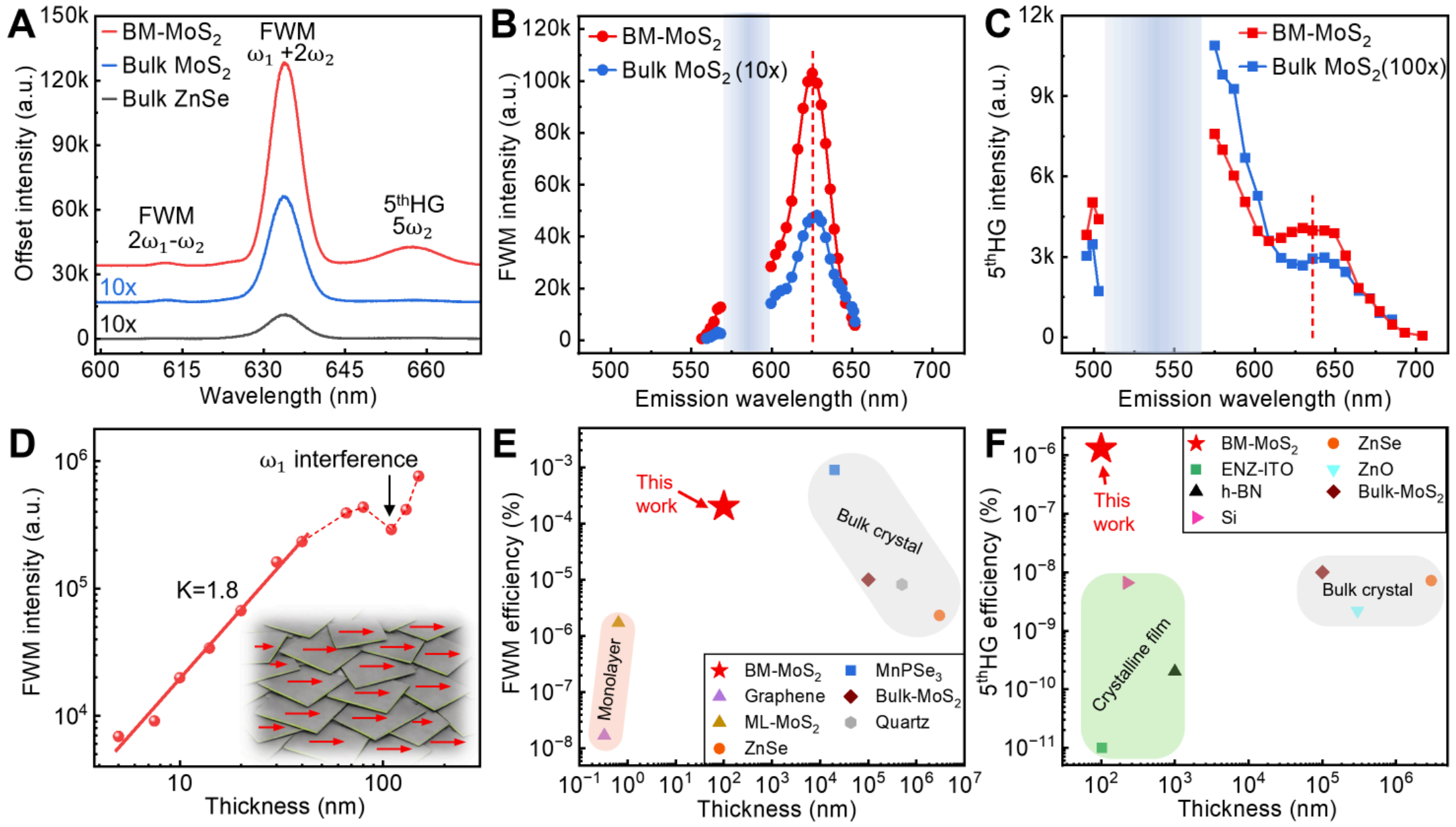

**Fig. 4. Colossal FWM and $5^{th}$HG efficiency.** (**A**) Nonlinear emission spectra under two-color ($\omega_1$ and $\omega_2$) excitation from the BM-$MoS_2$ thin film (~100 nm thick, red curve), pristine bulk $MoS_2$ crystal (~100 μm thick, blue curve, amplified by 10x), and bulk ZnSe crystal (~3 mm thick, black curve, amplified by 10x). (**B** and **C**) Extracted FWM (B) and $5^{th}$HG (C) intensities at varied emission wavelengths of the BM-$MoS_2$ thin film (red) and pristine bulk $MoS_2$ crystal (blue). Nonlinear emission at mid-IR pump wavelengths that overlap with the water-vapor absorption band is omitted. (**D**) FWM intensity as a function of the film thickness. The solid line represents the power-law fits $T^K$. The dots represent the experimental results. Inset in (D): the nonlinear field generated by each monolayer $MoS_2$ flake is coherently polarized in the same direction of the pump laser, so that the light field adds up constructively. (**E** and **F**) Comparisons of conversion efficiencies of FWM (**E**) and non-cascaded $5^{th}$HG (**F**) in different types of materials, including graphene (*42*), monolayer $MoS_2$ (ML-$MoS_2$) (*36*), bulk ZnSe crystal (measured in this work), bulk $MnPSe_3$ crystal (*14*), bulk $MoS_2$ crystal (measured in this work), bulk quartz crystal (*14*), bulk ZnO crystal (*43*), BM-$MoS_2$ thin film (measured in this work), epsilon-near-zero indium tin oxide thin film (ENZ-ITO) (*44*), Si thin film (*15*), twisted h-BN thin film (*16*). The conversion efficiencies among different materials are extracted at the same pumping power densities as our studies ($\omega_1$: 30 GW $cm^{-2}$; $\omega_2$: 56 GW $cm^{-2}$). See details in Table S1.

Compared to bulk $MoS_2$ crystals, both FWM and $5^{th}$HG signals in BM-$MoS_2$ show their strongest enhancement near resonant emission wavelengths (red vs. blue curves in Fig. 4B and C). The BM-$MoS_2$ thin film exhibits **a striking 152-fold enhancement in $5^{th}$HG at 630** nm over the bulk $MoS_2$ crystals of 100 $\mu$m thick (Fig. 4C), whereas the highest enhancement of FWM reaches

~22-fold at 625 nm (Fig. 4B). At off-resonant wavelengths, both FWM and 5$^{th}$HG efficiencies decrease substantially (Fig. 4B and C). **These observations highlight that the excitonic resonances can substantially amplify the nonlinear transition dipole moment across the optically active region (*45, 46*).** The BM-$MoS_2$ resonance is only slightly blueshifted (~10 meV) relative to bulk $MoS_2$. This behavior is consistent with the small blueshift of the monolayer A-exciton resonance compared to bulk $MoS_2$ that has been reported in linear absorption studies. In the monolayer limit, the electronic band gap increases owing to the absence of interlayer coupling. At the same time, however, the exciton binding energy also increases substantially due to reduced dielectric screening. These two effects partially compensate each other, resulting in an A-exciton resonance in monolayer $MoS_2$ that remains close in energy to that of bulk $MoS_2$ (*47*).

Beyond excitonic effects, the unique superlattice architecture of BM-$MoS_2$ can further strengthen its giant 5$^{th}$HG response in the non-perturbative regime. As illustrated schematically in Fig. 3F, non-perturbative HHG proceeds through field-induced interband tunneling that generates free electron-hole pairs, followed by their acceleration and recollision, which produces the emitted high-harmonic photons. In BM-$MoS_2$, the insulating, low-dielectric-constant PVP interlayer spacer not only confines photo-carriers within a highly two-dimensional (2D) space but also strengthens electron-hole Coulomb interactions, which may substantially boost the probability of electron-hole recollision that are fundamentally responsible for nonperturbative HHG in solids (*22*).

Nonlinear optical responses are often sensitive to crystal orientation, generally requiring single-crystal nonlinear materials for practical NLG. The solution-assembled BM-$MoS_2$ consists of monolayer flakes that are randomly oriented in-plane (Fig. S6), which could hinder a fully constructive buildup of the nonlinear fields in certain circumstances. Nonetheless, **monolayer $MoS_2$ possesses $D_{3h}$ symmetry, which renders odd-order nonlinear processes such as FWM and 5$^{th}$HG insensitive to in-plane crystal orientation** (Supplementary Note 1) (*22, 48*). This orientation independence ensures that **nonlinear fields emitted from different flakes are polarization-matched and thus add up constructively (see inset of Fig. 4D), leading to an expected quadratic scaling of NLG intensity with the film-thickness and hence the conversion efficiency**.

To experimentally verify this prediction, we measured the FWM intensity as a function of the BM-$MoS_2$ film thicknesses. Below a thickness of ~ 60 nm, the FWM intensity scales with the film thickness following a power-law exponent K=1.8 (Fig. 4D), approaching the quadratic dependence expected for coherent field addition and confirming coherent buildup of nonlinear fields across stacked monolayers. As the thickness increases beyond 80 nm, the FWM intensity exhibits a slight drop, followed by a rapid rise. This non-monotonic behavior is attributed to the interference of the pump beam due to multiple internal reflections, i.e. a cavity effect, as confirmed by our numerical simulations (Fig. S7).

We also note that earlier studies reported that FWM output from pristine TMD multilayers typically scales linearly or sub-linearly with thickness and rapidly saturates at a just few nanometers (*38*). As the layer number increases beyond the monolayer, the interlayer electronic coupling in pristine TMDs significantly perturbs the electronic band structures, driving a transition from a direct to an indirect bandgap, and substantially increasing dielectric screening, all of which weaken exciton resonances and reduce the nonlinear susceptibility. In contrast, the **BM-$MoS_2$ exhibits the inherent monolayer-like optical nonlinearity while simultaneously offering extended light-matter interaction length and preserving the coherent nature of emitted signals in the subwavelength regime.** This combination enables BM-$MoS_2$ to sustain efficient, constructive buildup of nonlinear light emission across the volume under pumping, contributing to the exceptional conversion efficiency. Importantly, the phase-matching constraints, which typically limit the ultimate efficiency of nonlinear crystals with macroscopic dimensions, are largely relaxed in BM-$MoS_2$ thin films. The coherence length ($L_c$) for pristine $MoS_2$ is estimated to be ~ 340 nm from $L_c = \frac{\pi}{|\Delta k|}$, where $\Delta k$ denotes the wavevector mismatch between the driving fields and the generated field for this FWM process (*1*). BM-$MoS_2$ is in principle expected to exhibit an even longer $L_c$ than pristine $MoS_2$ because the interlayer PVP is less dispersive than $MoS_2$. In this case, **the BM-$MoS_2$ film thicknesses investigated here remain far below $L_c$, allowing the coherent buildup of the nonlinear fields without the stringent phase-matching requirements imposed on conventional bulk materials.**

We have also examined the thickness dependence of SFG ($\omega_1 + \omega_2$), which is driven by the same pair of pump beams used for FWM but behaves differently. Intriguingly, we observed a linear thickness dependence of SFG (Fig. S8). Distinct from the FWM originating from $\chi^{(3)}$ nonlinearity, SFG belongs to $\chi^{(2)}$ nonlinearity and its polarization is modulated by the in-plane orientation of each flake (Supplementary note 1) (*49*). As a result, the vectorial summation of SFG fields from different monolayers can be regarded as a 2D random walk scheme, producing a net field intensity that scales linearly with the film thickness (see details in Supplementary Note 2).

We further performed polarization-resolved measurements to characterize the polarization state of the nonlinear radiation in our experimental geometry (see details in Fig. S9A and B). Our measurements reveal that the SHG signal is only partially polarized with a relatively low degree of polarization ($DOP = (I_{max} - I_{min})/(I_{max} + I_{min})$ ≈0.5) (Fig. S9C). Since the pump spot in our measurement is relatively large (~50 μm) compared with the typical lateral size of an individual monolayer size (~1-2 μm), the measured SHG is effectively an ensemble/spatial average over multiple monolayer flakes with different crystal orientations. Because different nanosheets exhibit different nonlinear polarization orientations, spatial averaging over the ensemble naturally reduces the measured net polarization contrast. Consistently, rotational-anisotropy SHG (RA-SHG) measurements on the BM-$MoS_2$ film reveal a nearly circular angular dependence, suggesting that the characteristic six-lobed anisotropy of individual monolayer $MoS_2$ is largely averaged out by the orientational distribution of nanosheets within the large pump spot (see details in Fig. S10). In contrast, the odd-order nonlinear response (FWM and $5^{th}$HG) of monolayer $MoS_2$ is effectively independent of the flake orientation and polarized in the same direction of the pump lasers, as discussed in Supplementary Note 1. This behavior is consistent with our observation that both $5^{th}$HG and FWM signals are nearly fully polarized, with similar $DOP$ values of ~0.96 (Fig. S9D and E).

The strong NLG allows direct measurements of conversion efficiencies using a calibrated power meter. Under two-color excitation, the FWM conversion efficiency of the BM-$MoS_2$ thin film is estimated to be ~ $2\times10^{-4}$ % (see Methods for details). Fig. 4E compares the FWM efficiencies of the BM-$MoS_2$ with other nonlinear materials of varied thicknesses under a comparable pumping power density. The conversion efficiency of pristine monolayer $MoS_2$ is two orders of magnitude lower than that of the BM-$MoS_2$, primarily attributed to its atomically thin optical length (*36*). Compared to monolayer $MoS_2$, graphene exhibits an even lower FWM efficiency of ~ $10^{-8}$ %, owing to its gapless electronic structure and the absence of exciton resonance enhancement (*42*). Furthermore, the FWM output of the BM-$MoS_2$ substantially outperforms that of most conventional nonlinear crystals such as quartz and ZnSe, whose thicknesses far exceed the pump wavelength. When phase matching is not satisfied, the generated field in macroscopic crystals shifts its phase from the pump field over wavelength-scale distances, preventing coherent buildup of the FWM signal over the propagation length.

Under single-color excitation ($\omega_2$ = 56 GW cm$^{-2}$), the conversion efficiency of the 5$^{th}$HG in the BM-$MoS_2$ thin film reaches about $1.3\times10^{-6}$ % (see Methods for details). Notably, **this 5$^{th}$HG efficiency surpasses most previously reported THG efficiencies in pristine TMDs by two to three orders of magnitude** (Table S1), despite being a higher harmonic order. We further note that conversion efficiencies beyond the 3$^{rd}$ harmonic have been rarely reported in TMDs, underscoring the exceptional HHG response enabled by the BM-$MoS_2$. Fig. 4F benchmarks the conversion efficiencies of non-cascaded 5$^{th}$HG across different materials, where we extracted reported efficiencies at the same pump power density used in our study (56 GW cm$^{-2}$) for a fair assessment. Strikingly, **the 5$^{th}$HG efficiency of the BM-$MoS_2$ thin film exceeds that of conventional bulk crystals such as Si, ZnO and ZnSe by 2 to 3 orders of magnitude (*15, 43*)**. Moreover, it even surpasses the purpose-engineered meta-photonic platforms such as Si metasurfaces and is comparable to that of an epsilon-near-zero (ENZ) material (In:CdO) embedded in an optical cavity (*10, 15*). Such meta-structures can support resonant modes for strong local field enhancement and have been regarded as among the most efficient solid-state systems for HHG (*10, 15, 19*). However, their operation is constrained to a narrow spectral window (typically<30 nm) set by the resonance mode (*10, 19, 50*). In contrast, HHG in BM-$MoS_2$ arises from the intrinsic nonlinear polarization of the material rather than resonant photonic modes, allowing it to achieve high 5$^{th}$HG efficiency across a broad spectral range (>100 nm) (Fig. 4C). We note that the conversion efficiency is not limited by sample damage as the BM-$MoS_2$ thin film remained stable under the present measurement conditions. Based on the measured power law, the 5$^{th}$HG and FWM efficiencies are expected to further increase with the enhancement of the pump power density. A systematic high-fluence and damage-threshold study will be an important direction for future work, particularly to determine how the PVP interlayer encapsulation in BM-$MoS_2$ influences the optical damage threshold.

We further investigated the roles of phase matching and self-absorption in limiting the nonlinear conversion efficiencies through transfer-matrix-method (TMM) simulations of thickness-dependent FWM and 5$^{th}$HG in BM-$MoS_2$ (see details in Supplementary Note 3). Our simulation results indicate that self-absorption contributes to the efficiency loss, while phase mismatch becomes increasingly important as the BM-$MoS_2$ thickness is extended into the micrometer regime. Nevertheless, the BM-$MoS_2$ thin films investigated in this work are in the subwavelength thin-film regime, and thus the stringent phase-matching requirement that limits macroscopic bulk nonlinear crystals is largely relaxed. For substantially thicker BM-$MoS_2$ films, however, phase mismatch could become increasingly important, ultimately limiting further efficiency scaling with film thickness.

**Mid-IR-to-visible upconversion for molecular vibrational sensing**

With its colossal FWM efficiency, the BM-$MoS_2$ can function as a mid-IR-to-visible frequency upconverter. Specifically, the FWM process ($\omega_1 + 2\omega_2$) converts the mid-IR signal ($\tilde{\nu}_{mid-IR}$) into a visible idler emission ($\tilde{\nu}_{vis}$) following $\tilde{\nu}_{mid-IR}$=($\tilde{\nu}_{vis} - \tilde{\nu}_{pump}$)/2, where $\tilde{\nu}_{pump}$ corresponds to the wavenumber of the near-IR pump fixed at 1028.9 nm. Importantly, the FWM intensity largely remains within the same order of magnitude across our mid-IR frequency tuning range, indicating **a broad spectral tunability over 1000 nm in the mid-IR wavelength range** (Fig. 3B). When tuning the mid-IR pump wavelength through the water vapor absorption band (3500 to 4000 $cm^{-1}$), fine vibrational absorption features of water molecules imprint onto the mid-IR pump as it propagates through air, which are subsequently transduced onto the visible FWM spectra (Fig. 5A). Considering the limited IR detection technologies based on low-band gap materials, this mid-IR-to-visible transduction provides a powerful detection strategy, enabling molecular mid-IR vibrational features to be read out using standard, low-noise silicon photodetectors and compact visible-range spectrometers (*51, 52*).

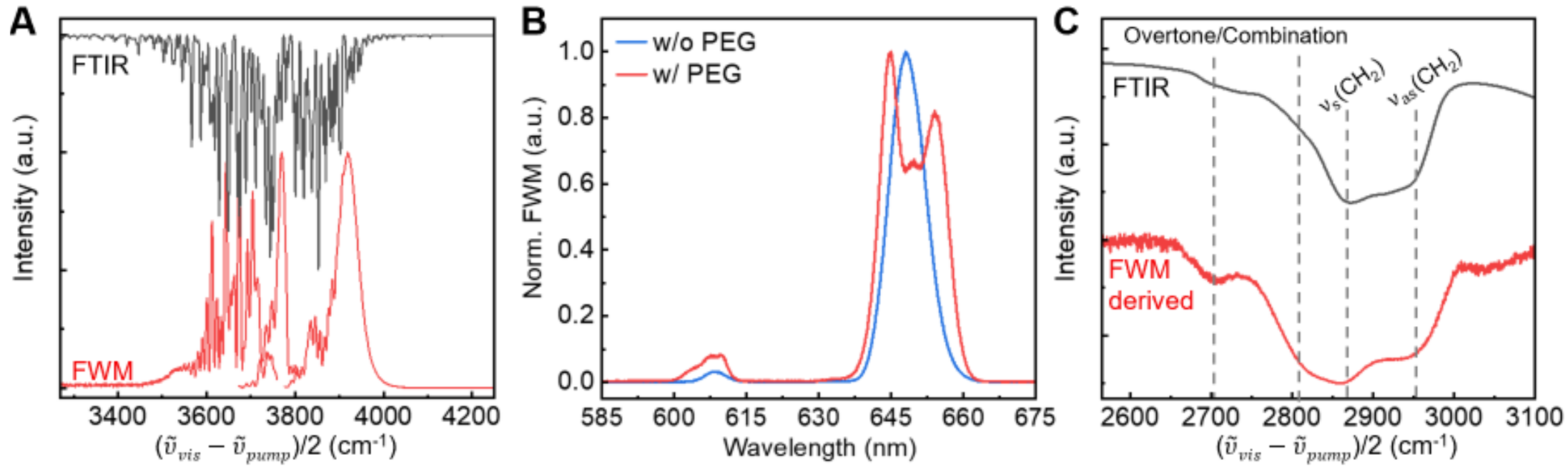

**Fig. 5. Mid-IR spectroscopy via FWM transduction.** (**A**) Mid-IR absorption of ambient water vapor resolved through the FWM transduction of the BM-$MoS_2$ (red). The water-vapor transmission spectrum obtained through FTIR measurements is displayed for reference (black). (**B**) FWM emission spectra collected from a BM-$MoS_2$ thin film with (red) and without (blue) a thin layer of polyethylene glycol (PEG) on top. (**C**) Mid-IR transmission spectrum of PEG extracted from the FWM measurements by taking the ratio of the two spectra in (B) and applying a square-root transformation to account for the power-law nature of the FWM process (red). The FTIR transmission spectrum of PEG (black) is displayed for reference.

Beyond gas-phase detection, we further demonstrated the vibrational sensing in condensed matter. A thin layer of polyethylene glycol (PEG) was drop-cast onto the BM-$MoS_2$ thin film, and the FWM emission spectra with and without PEG were used to retrieve the mid-IR vibrational absorption (Fig. 5B). The pump lasers were attenuated to avoid sample burning and thus FWM operates in the perturbative regime. In the presence of PEG, two dips emerge on the FWM spectrum, in agreement with the characteristic -$CH_2$- symmetric and asymmetric vibrational modes of PEG (Fig. 5B). Specifically, the PEG-induced mid-IR attenuation is obtained by dividing the two FWM spectra in Fig. 5B and taking the square root, $T_{PEG} \propto \sqrt{\frac{I_{FWM,PEG}}{I_{FWM,bare}}}$, which is expected for the power law of the FWM process on the mid-IR intensity. The resulting spectrum (red curve, Fig. 5C) clearly resolves the characteristic -$CH_2$- symmetric and asymmetric stretching modes and the overtone and combination modes of PEG, in agreement with the FTIR measurements (black curve, Fig. 5C). These results showcase BM-$MoS_2$ as a broadband, high-sensitivity, room-temperature platform for mid-IR vibrational detection via visible-band readout.

## DISCUSSION

In summary, we demonstrate colossal high-order nonlinear optical processes in solution-processed BM-$MoS_2$ thin films. The unique "bulk monolayer" architecture preserves the exceptional optical nonlinearity of monolayer $MoS_2$ while providing a thickness-scalable nonlinear light generation. Under moderate pumping (<0.1 TW $cm^{-2}$), a 100 nm thick BM-$MoS_2$ thin film delivers an extraordinary power conversion efficiency of $2\times10^{-4}$ % for FWM and $1.3\times10^{-6}$ % for 5$^{th}$HG over a broad spectral regime, greatly outperforming the conventional nonlinear crystals despite being orders of magnitude thinner. The strong 5$^{th}$HG and FWM in BM-$MoS_2$ mainly arise from two factors. **First, monolayer semiconducting TMDs intrinsically possess large optical nonlinearities.** Excitonic resonances can substantially amplify the nonlinear transition dipole moment across the optically active spectral region, as supported by our wavelength-dependent excitation spectroscopy. For 5$^{th}$HG in the non-perturbative regime, HHG proceeds through field-induced interband tunneling and electron-hole recollision (Fig. 3F). Confining photo-carriers within a highly 2D space may strengthen electron-hole Coulomb interactions, which may also substantially boost the probability of electron-hole recollision that are fundamentally responsible for non-perturbative HHG in solids. **Second, the BM-$MoS_2$ architecture overcomes the atomically thin interaction-length limitation of isolated monolayers.** The emitted nonlinear fields can build up coherently across the stacked monolayers in the subwavelength regime. Furthermore, the subwavelength-thick BM-$MoS_2$ films remain within the relevant coherence length, so the stringent phase-matching constraints that typically limit macroscopic nonlinear crystals are largely relaxed.

A recent demonstration highlights the exciting possibility of integrating vdW materials with high-quality photonic circuitry for enhanced nonlinear optical processes (*53*). In this context, BM-$MoS_2$ thin films may provide a complementary scalable material platform for future integrated nonlinear photonics. Because BM-$MoS_2$ combines strong high-order nonlinear response, broadband mid-IR-to-visible frequency conversion, and versatile thin-film processability, integrating it with engineered photonic structures such as waveguides, resonators, or microcavities could further enhance the local optical field, improve outcoupling, and enable compact on-chip high-order nonlinear light generation. More broadly, this thin-film platform may also open a pathway toward attosecond pulse generation in ultra-compact device architectures and extend coherent light generation toward the extreme-ultraviolet regime. Our study establishes a scalable, low-cost, integrable, and high-performance nonlinear thin film material platform, which can drive advances in on-chip integrated devices to the regime of extreme nonlinear optics.

## MATERIALS AND METHODS

**Preparation of the $MoS_2$ ink and BM-$MoS_2$ thin films.** We used polymer-assisted electrochemical intercalation and sonication processes to prepare the ink. For electrochemical intercalation, a thin piece of $MoS_2$ bulk crystal was attached to a copper clamp as the cathode and a graphite rod was used as the anode. Tetraethylammonium bromide (THAB, 98% from TCI) and PVP (Sigma-Aldrich, molecular weight ~ 40,000) were dissolved in dimethylformamide (DMF) solution as the electrolyte with a concentration of 5 and 20 mg/mL, respectively. After dipping the cathode and anode into the electrolyte, a high voltage of -8.5 V was applied to the $MoS_2$ cathode during the electrochemical intercalation, extended over ~150 minutes. During this step, $THA^+$ cations and PVP intercalate into the van der Waals gap of $MoS_2$, resulting in a great expansion of the bulk crystal. Afterward, the intercalated crystal was immediately subject to sonication in a PVP/DMF solution (18 mg/mL) for ~ 8 minutes, creating an ink consisting of exfoliated $MoS_2$ nanosheets adhered with PVP ligands. High-speed centrifugation at 12100 rpm was then used to exchange the DMF solvent into IPA, followed by two IPA washes through centrifugation. We iteratively centrifuged the ink at 3500 rpm for 5 minutes and discarded the precipitates to obtain a monolayer $MoS_2$ dispersion where thick nanosheets were fully removed from the ink. Finally, we used the ink material to prepare BM-$MoS_2$ thin films via spray-coating. The BM-$MoS_2$ thin film was prepared by spray coating the ink on substrates (e.g., Si wafers) at room temperature under ambient conditions. Specifically, the sprayer was connected to a nitrogen tank, with a pressure regulator set to ~5 psi above atmospheric pressure for controlled spraying. The nozzle-to-substrate distance was kept at around 1 cm throughout the process. The film thickness is tuned by the spraying time.

**Structural characterizations.** XRD characterizations were performed at the BioPACIFIC MIP user facilities at California NanoSystem Institute (CNSI), UC Santa Barbara. The instrument was custom-built using a high brightness Ga alloy liquid metal jet X-ray source (Excillum MetalJet D2+ 70 keV, 9.24 keV) and a 4-megapixel hybrid photon-counting X-ray area detector (Dectris Eiger2 R 4M). Other structural characterizations were performed using HRTEM (Titan S/TEM (FEI): 300 kV acceleration voltage), and AFM (Bruker Dimension Icon Scanning Probe Microscope) at CNSI, UC Los Angeles.

**Optical characterizations.** Nonlinear measurements were performed using a 45° reflection geometry using a home-built system under p-polarized pumping. The nonlinear signal is dispersed by a spectrograph (Shamrock, Andor) and detected by a charge-coupled device (CCD) (Newton idus, Andor). The Raman and PL spectra were carried out using a Horiba Lab Ram 401 HR Evolution confocal Raman system. Pristine bulk $MoS_2$ was measured using a longer integration time due to its weak signal. All PL intensities were therefore normalized to their respective integration times before comparison. For the $5^{th}$HG measurement, the BM-$MoS_2$ thin film was pumped by a mid-IR laser generated from an optical parametric amplifier (Light Conversion, ORPHEUS-ONE-HP) with an average power of 264 mW, a repetition rate of 100 kHz, a pulse duration of 150 fs, and a Gaussian beam diameter of 200 μm defined by a knife-edge experiment at the $1/e^2$ intensity level. The generated $5\omega_2$ signal at 638.0 nm under 3190 nm pump was measured using a calibrated power meter after spectral filtering, giving an average output power of approximately 3.4 nW and the $5^{th}$HG conversion efficiency can be determined by the ratio of the generated $5^{th}$HG output power to the incident pump power: $\eta_{5\omega_2} = \frac{P_{5\omega_2}}{P_{\omega_2}} = \frac{3.4\, nW}{264\, mW} \approx 1.3 \times 10^{-6}\%$. For the FWM measurements, the BM-$MoS_2$ thin film was excited by both the near-IR beam (Pharos) with an average power of 118 mW, a repetition rate of 100 kHz, a pulse duration of 2 ps, and a $1/e^2$ beam diameter of 50 μm; and the same mid-IR beam ($\omega_2$) used for $5^{th}$HG measurement, with an average power of 264 mW, a repetition rate of 100 kHz, a pulse duration of 150 fs, and a $1/e^2$ beam diameter of 200 μm. Because the FWM process occurs only within the common spatiotemporal overlap of the two pump beams, we estimated the effective overlap-corrected pump powers as follows. Taking the smaller near-IR beam footprint (50 μm) as the effective spatial-overlap region, the fraction of the much larger mid-IR beam (200 μm) power within this region is estimated as $P_{\omega_2,overlap} = 264\, mW \times \left(1 - e^{-2\left(\frac{50}{200}\right)^2}\right) \approx 31\, mW$. Temporally, only the 150-fs portion of the 2 ps near-IR pulse overlaps with the mid-IR pulse and thus the temporal-overlap fraction for the near-IR beam is estimated as $P_{\omega_1,overlap} = 118\, mW \times \left(\frac{150\, fs}{2000\, fs}\right) \approx 8.9\, mW$. The measured output power after spectral filtering at the FWM wavelength was ~79.8 nW and thus we can estimate the FWM conversion efficiency by taking the ratio of the FWM output power and the spatiotemporal overlapped pump powers: $\eta_{FWM} = \frac{P_{FWM}}{P_{\omega_1,overlap} + P_{\omega_2,overlap}} = \frac{79.8\, nW}{8.9\, mW + 31\, mW} \approx 2.0 \times 10^{-4}\%$.

## SUPPLEMENTARY MATERIALS

Figs. S1 to S11

Supplemental note 1 to 3

Table S1

References

**ACKNOWLEDGEMENT:** The authors acknowledge Electron Imaging Center for NanoMachines (EICN) and Nano&Pico Characterization Lab (NPC) at California NanoSystems Institute (CNSI). X.D. acknowledges the support from the Office of Naval Research through grant no. N00014-22-1-2631. W.X. acknowledges the Brown Institute for Basic Science (award no. S643177) and the Office of Naval Research (N000142412262). The core-loss EELS experiments were performed using facilities and instrumentation at the UC Irvine Materials Research Institute (IMRI), which is supported in part by the National Science Foundation through the UC Irvine Materials Research Science and Engineering Center (DMR-2011967) (for X.P.).

**AUTHOR CONTRIBUTIONS:** X.D., W.X. and Y.H. conceived and supervised the research. B.Z. developed the material design and synthesis, and conducted the PL, Raman, AFM measurements, and optical simulations. Y.J. and C.C.Y. performed FWM and 5$^{th}$HG nonlinear measurements with the assistance of B.Z. H.L. contributed material preparation and assisted with optical measurements. R.W. performed SEM and TEM characterizations. X.Y. and X.P. contributed to EELS and STEM measurements. B.Z. and X.D. co-wrote the manuscript with input from all authors.

**COMPETING INTERESTS:** The authors declare no competing interests.

**DATA, CODE, AND MATERIALS AVAILABILITY:** All data and code needed to evaluate and reproduce the results in the paper are present in the paper and/or Supplementary Materials. This study did not generate new materials.